\documentclass{article}

\usepackage[utf8]{inputenc} 
\usepackage[T1]{fontenc}    
\usepackage{hyperref}       
\usepackage{url}            
\usepackage{booktabs}       
\usepackage{amsfonts}       
\usepackage{microtype}      
\usepackage{lipsum}
\usepackage{fancyhdr}       
\usepackage{graphicx}       
\usepackage{siunitx}
\usepackage{amsmath}
\usepackage{multirow}
\newcommand{\LP}[2][]{$\mathrm{LP}_{#2}^{#1}$}
\newcommand{\LPbf}[2][]{$\mathbf{LP}_{\bf #2}^{\bf #1}$}
\usepackage{float}

\title{Multiple intermodal-vectorial four-wave-mixing bands generated by selective excitation of orthogonally polarized \LPbf{01} and \LPbf{11} modes in a birefringent fiber}

\author{
  Sylwia Majchrowska, Kinga Żołnacz, Wacław Urbańczyk, Karol Tarnowski \\
  Wroclaw University of Science and Technology,  \\
  Faculty of Fundamental Problems of Technology, \\
  Department of Optics and Photonics, \\
  Wybrzeże Wyspiańskiego 27, 50-370 Wrocław, Poland\\
  \texttt{\{sylwia.majchrowska, karol.tarnowski\}@pwr.edu.pl} \\
}

\begin{document}
\maketitle

\begin{abstract}
This study investigated the nonlinear frequency conversions between the six polarization modes of a two-mode birefringent fiber.
The aim was to demonstrate that the selective excitation of different combinations of linearly polarized spatial modes at the pump wavelength initiates distinct intermodal vector four-wave mixing processes.
In particular, this study shows that exciting two orthogonally polarized \LP{01} and \LP{11} modes can result in simultaneous generation of up to three pairs of different spatial modes of orthogonal polarizations at different wavelengths.
The role of the phase birefringence of the spatial modes in the phase-matching of such a four-wave mixing process was explained.
Moreover, the theoretical predictions were verified through numerical simulations based on coupled nonlinear Schrodinger equations and also confirmed experimentally in a commercially available birefringent fiber.
\end{abstract}

\noindent The nonlinear phenomena in multimode and few-mode fibers have been extensively studied and continue to garner attention~\cite{Krupa2019}, particularly
for the following reasons: (i) expectations related to the application of spatial division multiplexing in telecommunication networks \cite{Winzer2018},
(ii) new possibilities for the observation of nonlinear phenomena provided via compact and powerful laser sources~\cite{Zervas2014}.
The newly observed phenomena are multimode solitons~\cite{Zhu2016, Zitelli2021},
beam self-cleaning \cite{Wright2016, Krupa2017},
and discretized conical emission \cite{Tarnowski2021, Kibler2021}.
In addition, other nonlinear phenomena such as a~Raman scattering~\cite{Rishoj2019} and a~four-wave mixing~\cite{Perret2019, Dupiol2017} have also been revisited.

Recent studies on intermodal four-wave mixing (IM-FWM) have focused on its impact on supercontinuum generation in a few-mode step-index fiber~\cite{Perret2019},
the modulation instability in a graded-index fiber~\cite{Dupiol2017},
and the proof-of-concept application of IM-FWM to mode-wavelength conversion in mode-multiplexing telecommunication schemes~\cite{Friis2016}.
These studies reported the observance of IM-FWM involving two spatial modes of a nonbirefringent fiber.
Nonlinear conversion occurs from two excited modes to signal and idler bands, each generated in a different mode from the set of excited modes.
For specific fiber and excitation conditions, the phase-matching condition and overlap coefficients determine (i) the modes that are involved in FWM, (ii) the positions of the generated bands, and (iii) the efficiency of conversion.
For example, IM-FWM was observed for the following pairs of modes: \LP{01}-\LP{02}, \LP{01}-\LP{11}, and \LP{11}-\LP{21}~\cite{Dupiol2017}.
Moreover, the conversion may also occur for other pairs involving higher-order spatial modes of nonbirefringent fibers~\cite{Perret2019}.
In birefringent fibers, the FWM involves polarization modes~\cite{Millot2014}.
In particular, phase-matching can be achieved between two polarization modes of a single (typically fundamental) spatial mode~\cite{Stolen1981}.
This type of vectorial FWM has been investigated in the context of sensing~\cite{Tarnowski2013}
and entangled photon pairs generation~\cite{Smith2009}.
However, in a few-mode birefringent fiber, more complex FWM processes involving intermodal-vectorial interactions are also possible.
Garay-Palmett et al.~\cite{Garay-Palmett2016} presented an experimental and theoretical analysis of this type of FWM.
In their experiments, pairs of polarization modes were excited (limited to a single polarization direction), and consequently, orthogonally polarized signal and idler bands were observed.

In this study, a variety of IM-FWM processes that occur in step-index birefringent fibers supporting \LP{01} and \LP{11} spatial modes were investigated.
Further, experimental differentiation of distinct IM-FWM processes was possible employing the recently described method of exciting different combinations of polarization/spatial modes using a Wollaston prism~\cite{Zolnacz2022}.
In particular, pairs of different spatial modes of orthogonal polarizations can be purely excited using this method. This enabled the experimental investigation of the intermodal-vectorial FWM processes in a two-mode birefringent fiber for the first time.
In contrast to the excitation of only one spatial mode of orthogonal polarizations (vectorial FWM) or two spatial modes of the same polarization (intermodal FWM), which results in the generation of one pair of signal and idler bands (in different polarizations or different spatial modes, respectively), the mixed intermodal-vectorial FWM process produces up to three pairs of signal/idler bands simultaneously at different spatial modes and orthogonal polarizations.
Further, this study explained that the difference in the phase modal birefringence in the \LP{01} and \LP{11} spatial modes is responsible for the observed multiplication of the sidebands.

The phase-matching condition for the FWM process can be expressed as $\beta^{(l)} + \beta^{(m)} = \beta^{(p)} + \beta^{(n)}$,
where $\beta^{(i)}$ is the wave vector of the mode $i$.
Here superscripts $l$~and~$m$ were used to indicate the pump modes, while superscripts $p$~and~$n$~denoted the signal and idler modes, respectively.
Assuming that both excited modes were at a single frequency, and expanding the propagation constants in the Taylor series up to the second order at this frequency, the following relation is obtained:
\begin{multline}
   \frac{\beta^{(p)}_2 + \beta_2^{(n)}}{2}\Omega^2 + 
   \left(\beta_1^{(p)} - \beta_1^{(n)}\right)\Omega + \left(\beta_0^{(p)} - \beta_0^{(l)} -
    \beta_0^{(m)} + \beta_0^{(n)}\right) = 0,
    \label{eqn:phase-matching condition}
\end{multline}
where the subscripts indicate the derivative order, and the angular frequency detuning of a signal (idler) band is denoted by $\Omega$ ($-\Omega$).
Subsequently, this condition can be rewritten as:
\begin{equation}
   \bar\beta^{(p,n)}_2\Omega^2 + 
   \Delta\beta_1^{(p,n)}\Omega + \left(\Delta\beta_0^{(p,l)} - \Delta\beta_0^{(m,n)} \right) = 0,
    \label{eqn:phase-matching condition 2}
\end{equation}
where $\bar\beta^{(p,n)}_2$ is the mean chromatic dispersion of the signal/idler modes, $\Delta\beta_1^{(p,n)}$ corresponds to the difference in their group refractive indices at the pump wavelength, and $\Delta\beta_0^{(p,l)}$ and $\Delta\beta_0^{(m,n)}$ correspond to the difference in the phase refractive indices at the pump wavelength.

If only two modes are involved, like in the case of vectorial or intermodal FWM, $p = l$ and $m = n$, respectively.
Consequently, the free term in Eq.~\ref{eqn:phase-matching condition 2} vanishes and thereby yielding one trivial solution $\Omega = 0$ and one non-trivial solution $\Omega = -\Delta\beta_1^{(l,m)}\big/\bar\beta_2^{(l,m)}$.
In the general case of intermodal-vectorial FWM, the pump and generated band modes are of different orders and orthogonal polarizations.
Thus, to focus on this, it is assumed that the pump modes $(m,l)$ are the \LP[y]{01} and \LP[xe]{11} polarization modes.
In such a case, intermodal-vectorial FWM can occur in two different ways.
In the first scenario, which is always possible, the pump modes \LP[y]{01} and \LP[xe]{11} are converted to the same pair of modes and the spectral positions of the signal/idler bands are determined by the ratio of the difference in their group effective indices to the average dispersion, as previously explained.
In the second scenario, which is possible only if Eq.~(\ref{eqn:phase-matching condition 2}) has real solutions, four different modes are involved in the conversion process according to scheme $(m,l)\rightarrow(p,n)$. Thus, for a two-mode birefringent fiber and the assumed excitation, it is equivalent to the (\LP[y]{01},\LP[xe]{11})$\,\rightarrow\,$(\LP[x]{01},\LP[yo]{11}) process.
In such a case, the free term in
Eq.~\ref{eqn:phase-matching condition 2}, representing the difference in phase modal birefringence for the respective spatial modes, is not zero, resulting in the signal/idler bands being doubled.  

\vspace{10pt}
\begin{figure}[htbp]
\centering
\includegraphics[width=0.85\linewidth]{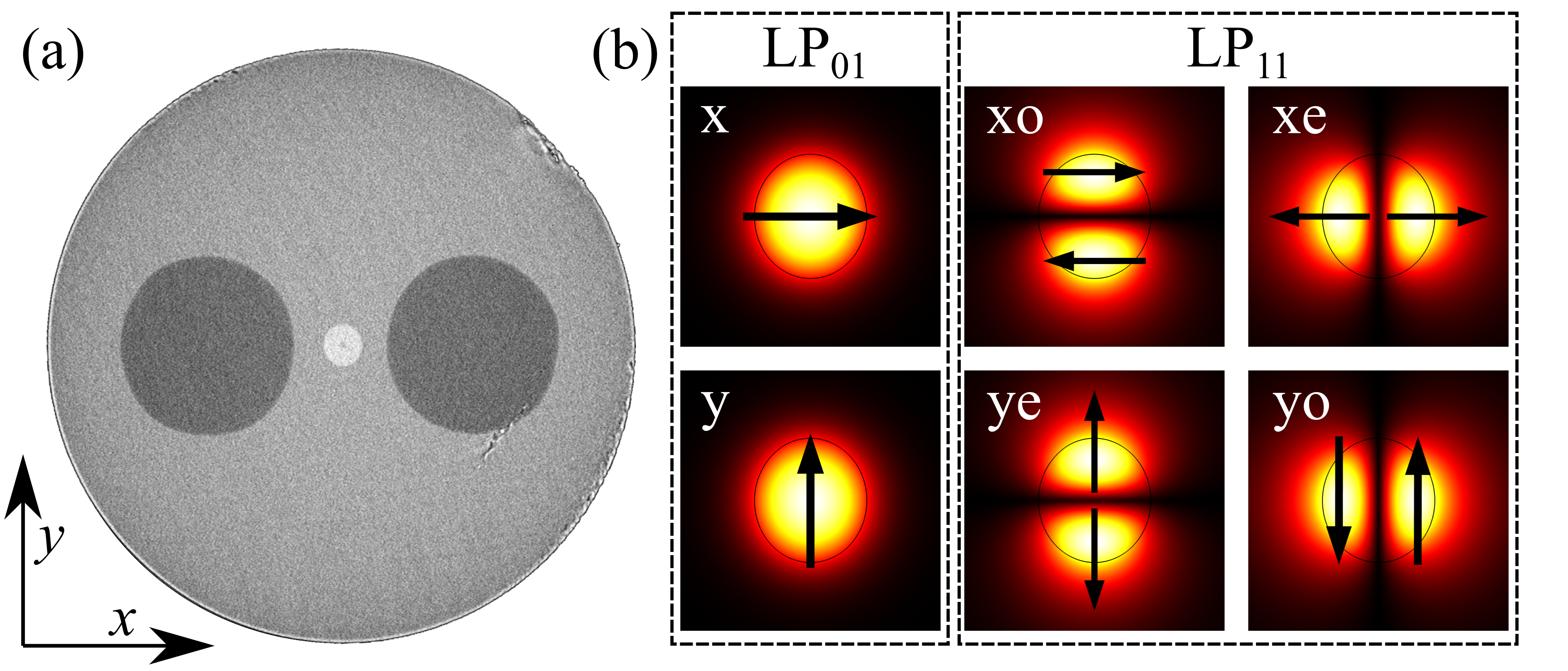}
\caption{(a) SEM image of the cross section of the birefringent fiber with stress-applying elements (Nufern PM1550B-XP). (b)~Electric field of the $x$- and $y$-polarized modes in the \LP{01} and \LP{11} groups.}
\label{fig:01_panda}
\end{figure}

However, phase-matching is not the only condition required for enabling intermodal FWM.
In addition, the mode overlapping coefficients $S^{(plmn)}_K$ and $S^{(plmn)}_R$ for the interacting modes are nonzero.
These coefficients appear in the coupled nonlinear Schr\"odinger equations (CNLSE)~\cite{Horak2012}:
\begin{align}
    \frac{\partial A^{(p)}}{\partial z} = & i\left(\beta_0^{(p)} - \beta_0^{(0)}\right)A^{(p)} - \left(\beta_1^{(p)} - \beta_1^{(0)}\right)\frac{\partial A^{(p)}}{\partial t}+\nonumber\\
    -&i\frac{\beta_2^{(p)}}{2}\frac{\partial^2 A^{(p)}}{\partial t^2} +i\frac{n_2\omega_0}{c}\left(1+\frac{i}{\omega_0}\frac{\partial}{\partial t}\right)\nonumber\\
    &\times\sum_{l,m,n}^{N-1}\left\{\left(1-f_R\right)S^{(plmn)}_KA^{(l)}A^{(m)}A^{(n)*} +\right.\nonumber\\ &\phantom{\times}\left.f_RS^{(plmn)}_RA^{(l)}\left[h*\left(A^{(m)}A^{(n)*}\right)\right]\right\},
    \label{eqn:CNLSE}
\end{align}
where $p$, $l$, $m$, and $n$ index $N$ modes are numbered from 0 to $N-1$, $A^{(i)}$ is the temporal amplitude envelope of mode $i$, $z$~is the propagation distance, $n_2$ is the nonlinear refractive index, $\omega_0$~is the central frequency, $f_R$ is the fractional Raman coefficient, $h$~is the Raman response function, and $*$ is an operator denoting convolution.
Overlap factors~\cite{Horak2012}:
\begin{subequations}
\label{eqn:overlaps}
\begin{align}
    S^{(plmn)}_R & = \textstyle\frac{\iint
    \left[\mathbf{F}_p^*\cdot\mathbf{F}_l\right]
    \left[\mathbf{F}_m\cdot\mathbf{F}_n^*\right]
    dxdy}
    {\sqrt{
    \iint\left|\mathbf{F}_p\right|^2
    dxdy
    \iint\left|\mathbf{F}_l\right|^2
    dxdy
    \iint\left|\mathbf{F}_m\right|^2
    dxdy
    \iint\left|\mathbf{F}_n\right|^2
    dxdy
    }},\label{eqn:SR}\\
    S^{(plmn)}_K & = \frac{2}{3}S^{(plmn)}_R + \nonumber\\ & +\frac{1}{3}
    \textstyle\frac{\iint
    \left[\mathbf{F}_p^*\cdot\mathbf{F}_n^*\right]
    \left[\mathbf{F}_m\cdot\mathbf{F}_l\right]
    dxdy}
    {\sqrt{
    \iint\left|\mathbf{F}_p\right|^2
    dxdy
    \iint\left|\mathbf{F}_l\right|^2
    dxdy
    \iint\left|\mathbf{F}_m\right|^2
    dxdy
    \iint\left|\mathbf{F}_n\right|^2
    dxdy
    }}\label{eqn:SK}
\end{align}
\end{subequations}
form a fourth-order tensor with $N^4$ elements ($N=6$ for the considered birefringent two-mode fiber).
However, most of the elements are zero because of mode symmetries~\cite{Poletti2009}.
The nonzero overlap coefficients correspond to the intermodal interactions that are allowed because of the selection rules~\cite{Poletti2009,Garay-Palmett2016}.

\begin{table}[b]
\centering
\caption{\bf Normalized overlap coefficients for intramodal and two-mode processes. The numerical values of the coefficients corresponding to the vectorial (intermodal) processes are $\underline{\mathrm{\bf underlined}}$ ($\overline{\mathrm{\bf overlined}}$).}
\small
\begin{tabular}{c|c||cc|c||cc|c}
    \multicolumn{2}{c||}{intramodal} & 
    \multicolumn{6}{c}{two-mode} \\
    \hline
    \hline
    $p$ & $f^{(pppp)}$ &
    $p$ & $n$ & $f^{(ppnn)}$ &
    $p$ & $n$ & $f^{(ppnn)}$ \\
    \hline
    $y$  &  1.000 & 
        $x$  & $y$  & $\underline{0.667}$ & 
            $x$  & $ye$ & $\underline{\overline{0.371}}$ \\
    $x$  &  1.000 & 
        $y$  & $ye$ & $\overline{0.557}$ & 
            $y$  & $xo$ & $\underline{\overline{0.371}}$ \\
    $xo$ &  0.794 & 
        $x$  & $xo$ & $\overline{0.557}$ & 
            $x$  & $yo$ & $\underline{\overline{0.354}}$ \\
    $ye$ &  0.791 & 
        $y$  & $yo$ & $\overline{0.530}$ &
            $y$  & $xe$ & $\underline{\overline{0.354}}$ \\
    $yo$ &  0.736 & 
        $x$  & $xe$ & $\overline{0.530}$ &
            $ye$ & $yo$ & $\overline{0.255}$ \\
    $xe$ &  0.731 & 
        $xo$ & $ye$ & $\underline{0.529}$ &
            $xo$ & $xe$ & $\overline{0.255}$ \\
         &        & 
         $xe$ & $yo$ & $\underline{0.490}$ &
            $xo$ & $yo$ & $\underline{\overline{0.170}}$ \\
         &        & 
            &        &        &
            $xe$ & $ye$ & $\underline{\overline{0.170}}$ \\
\end{tabular}
\label{tab:fplmn_SPM_XPM}
\end{table}

The nonzero overlap coefficients for the Nufern PM1550B-XP birefringent fiber used in the experiment were calculated.
The manufacturer specifications state the diameter and numerical aperture of the fiber core \SI{8.5}{\micro\meter} and numerical aperture \SI{0.125}{}, respectively.
The SEM image of the fiber cross-section and the normalized electric field of the \LP{01} and \LP{11} modes at pump wavelength \SI{1064.3}{\nano\meter} are shown in Figure~\ref{fig:01_panda}.
Moreover, the figure also explains the shortened notation ($x$, $y$, $xo$, $ye$, $xe$, and $yo$) used to indicate the respective polarization modes. 
The presented mode distributions calculated with the \makebox{COMSOL} Multiphysics mode solver were used to obtain the overlap factors using Eqs.~\ref{eqn:SR}~and~\ref{eqn:SK}.
The coefficient $S_K^{(x,x,x,x)}$ is the inverse of the effective mode area of the fundamental mode.
The magnitudes of the overlap factors were normalized to $S_K^{(x,x,x,x)}$; $f^{(plmn)} = \left|S_K^{(plmn)}\right|\big/S_K^{(x,x,x,x)}$ show the relative strength of nonlinear processes in reference to the intramodal nonlinearities of the fundamental mode~\cite{Poletti2009}: $f^{(pppp)}$ coefficients are used to describe intramodal nonlinearites of each mode, such as self-phase modulation, and $f^{(ppnn)}$ and $f^{(pnpn)}$ correspond to cross-phase modulation and two-mode FWM, respectively. The remaining coefficients govern other intermodal and/or vectorial FWM processes.

\begin{table}[t]
\centering
\caption{\bf Normalized overlap factors for selected intermodal-vectorial FWM. The relation $f^{(plmn)} = f^{(lpnm)}$ holds for all the combinations.}
\small
\begin{tabular}{cccc|c|cccc}
    $p$ & $l$ & $m$ & $n$ & $f^{(plmn)}$ &
    $p$ & $l$ & $m$ & $n$ \\
    \hline
    $ye$ & $y$  & $xo$ & $x$  & 0.370 & 
    $y$  & $ye$ & $x$  & $xo$ \\
    $yo$ & $y$  & $xe$ & $x$  & 0.353 &
    $y$  & $yo$ & $x$  & $xe$  \\
    $yo$ & $ye$ & $xe$ & $xo$  & 0.169 &
    $ye$ & $yo$ & $xo$ & $xe$  \\
\end{tabular}
\label{tab:fplmn_IMFWM}
\end{table}
\begin{table}[t]
\centering
\caption{\bf The measured modes' characteristics at \SI{1064.3}{\nano\meter}}
\small
\begin{tabular}{c|rrr}
    \multirow{2}{*}{$p$} & 
    \multicolumn{1}{c}{$D^{(p)}$} &
    \multicolumn{1}{c}{$\Delta N^{(p,y)}$} & 
    \multicolumn{1}{c}{\multirow{2}{*}{$\Delta n$}} \\
    & 
    \multicolumn{1}{c}{$\left[\frac{\si{\pico\second}}{\si{\kilo\meter\nano\meter}}\right]$} &
    \multicolumn{1}{c}{$\left[10^{-4}\right]$} & 
    \\
    \hline
    \hline
    $x$  & $-39.1$ &  $5.35$  & \multirow{2}{*}{$\Delta n^{(x,y)} = \SI{3.96e-4}{}$} \\
    $y$  & $-40.3$ &  $0.00$  & \\ \hline
    $xo$ & $-51.0$ & $10.77$  & 
    \multirow{2}{*}{$\Delta n^{(xo,ye)} = \SI{3.78e-4}{}$} \\
    $ye$ & $-51.2$ &  $5.33$  & \\ \hline
    $xe$ & $-52.0$ &  $8.51$  & \multirow{2}{*}{$\Delta n^{(xe,yo)} = \SI{4.20e-4}{}$} \\
    $yo$ & $-51.0$ &  $3.81$  &
\end{tabular}
\label{tab:fiber properties}
\end{table}

The normalized overlap factors for the intramodal and two-mode FWM presented in Table~\ref{tab:fplmn_SPM_XPM} indicate that the effective nonlinearities in the \LP{11} modes are lower than those in the \LP{01} modes.
In addition, the efficiency of the cross-polarization nonlinear process in a particular spatial mode is $2/3$ that of the intramodal nonlinearity.
Finally, two-mode FWM is possible even between two orthogonally polarized modes of different spatial groups.
The normalized overlap coefficients for all intermodal-vectorial FWM processes, that can occur by exciting orthogonally polarized modes of different spatial mode groups are presented in Table~\ref{tab:fplmn_IMFWM}.
There exist six such pairs; however, four of them can be excited within the applied excitation scheme~\cite{Zolnacz2022}: $(y,xo)$, $(x,ye)$, $(y,xe)$, and $(x,yo)$.
For such an FWM process, the free term in the phase-matching condition (Eq.~\ref{eqn:phase-matching condition 2}) is not zero, and two additional pairs of signal/idler bands are expected if phase-matching is fulfilled.

To predict the positions of the signal/idler bands generated by the intermodal, vectorial, and intermodal-vectorial FWM processes for the pump at \SI{1064.3}{\nano\meter}, the fiber was experimentally characterized.
Table~\ref{tab:fiber properties} summarizes the properties of the modes supported by the fibers.
The table lists the chromatic dispersion ($D^{(p)}$) of the mode $p$ measured with the white-light interferometry technique in a setup applying a spatial light modulator~\cite{Zolnacz2021}, the difference in group refractive indices ($\Delta N^{(p,y)}$) of the mode $p$ and mode $y$, and the difference in phase refractive indices of two polarization modes of a single spatial mode ($\Delta n$) measured using spectral interference with the lateral point force method~\cite{Kowal2018}.
Thus, knowing the processes that are allowed in terms of overlap factors, $\Omega$ was calculated by solving Eq.~\ref{eqn:phase-matching condition 2} for all cases. Consequently, the anticipated spectral positions of the signal and idler bands ($\lambda_\mathrm{t}^{(p)}$ and $\lambda_\mathrm{t}^{(n)}$), respectively, were obtained, as presented in Table~\ref{tab:FWM spectral}.

\begin{table}[t]
\centering
\caption{\bf Comparison of theoretical ($\lambda_\mathrm t [\si{\nano\meter}]$) and experimental ($\lambda_\mathrm e [\si{\nano\meter}]$) positions of signal/idler bands generated in different FWM from \SI{1064.3}{\nano\meter} pump; $l$ and $m$ denote two excited pump modes, $p$ and $n$ denote signal and idler modes; different types of FWM are indicated as V -- vectorial, I -- intermodal, IV -- intermodal-vectorial.}
\small
\begin{tabular}{c||r|r||r|r}
    process &
    \multicolumn{2}{c||}{theory} &
    \multicolumn{2}{c}{experiment} \\
    type:$(l,m)\rightarrow(p,n)$ &
    $\lambda^{(p)}_\mathrm{t}$ &
    $\lambda^{(n)}_\mathrm{t}$ &
    $\lambda^{(p)}_\mathrm{e}$ &
    $\lambda^{(n)}_\mathrm{e}$ \\\hline\hline
    V:$(x,y)\rightarrow(y,x)$ & 
        1021.2 & 1111.2 & 
        1020.5 & 1111.5 \\\hline 
    I:$(y,ye)\rightarrow(y,ye)$ & 
        1026.8 & 1104.6 &
        1027.4 & 1103.4 \\\hline 
    I:$(x,xo)\rightarrow(x,xo)$ & 
        1025.7 & 1106.0 &
        1024.6 & 1106.7 \\\hline 
    I:$(y,yo)\rightarrow(y,yo)$ & 
        1037.2 & 1092.9 &
        1038.1 & 1091.7 \\\hline 
    I:$(x,xe)\rightarrow(x,xe)$ & 
        1041.7 & 1087.9 &
        1040.9 & 1088.5 \\\hline 
    V:$(xo,ye)\rightarrow(ye,xo)$ & 
        1030.0 & 1101.0 & 
        1030.9 & 1099.7 \\\hline 
    V:$(xe,yo)\rightarrow(yo,xe)$ & 
        1034.7 & 1095.6 & 
        1036.1 & 1094.1 \\\hline 
    IV:$(x,ye)\rightarrow(ye,x)$ & 
        1064.3 & 1064.5 &
        1056.4 & 1072.5 \\ 
    IV:$(x,ye)\rightarrow(xo,y)$ & 
        1049.6 & 1079.4 &
        1048.8 & 1080.1 \\ 
    IV:$(x,ye)\rightarrow(y,xo)$ & 
        978.3 & 1166.9 & 
        978.0 & 1167.4 \\\hline 
    IV:$(y,xo)\rightarrow(y,xo)$ & 
         991.1 & 1149.2 &
         991.5 & 1148.5 \\ 
    IV:$(y,xo)\rightarrow(x,ye)$ & 
        \multicolumn{4}{c}{no phase-matching} \\
    IV:$(y,xo)\rightarrow(ye,x)$ & 
        \multicolumn{4}{c}{no phase-matching} \\\hline 
    IV:$(x,yo)\rightarrow(yo,x)$ & 
        1053.0 & 1075.8 & 
        1046.3 & 1083.9 \\ 
    IV:$(x,yo)\rightarrow(y,xe)$ & 
        \multicolumn{4}{c}{no phase-matching} \\
    IV:$(x,yo)\rightarrow(xe,y)$ & 
        \multicolumn{4}{c}{no phase-matching} \\\hline 
    IV:$(y,xe)\rightarrow(y,xe)$ & 
        1006.2 & 1129.5 &
        1005.9 & --- \\ 
    IV:$(y,xe)\rightarrow(x,yo)$ & 
        1027.5 & 1103.8 & 
        1026.2 & 1105.1  \\ 
    IV:$(y,xe)\rightarrow(yo,x)$ & 
        1016.9 & 1116.3 &
        1016.3 & 1116.5 \\ 
        
\end{tabular}
\label{tab:FWM spectral}
\end{table}

\begin{figure}[!ht]
\centering
\includegraphics[width=\linewidth]{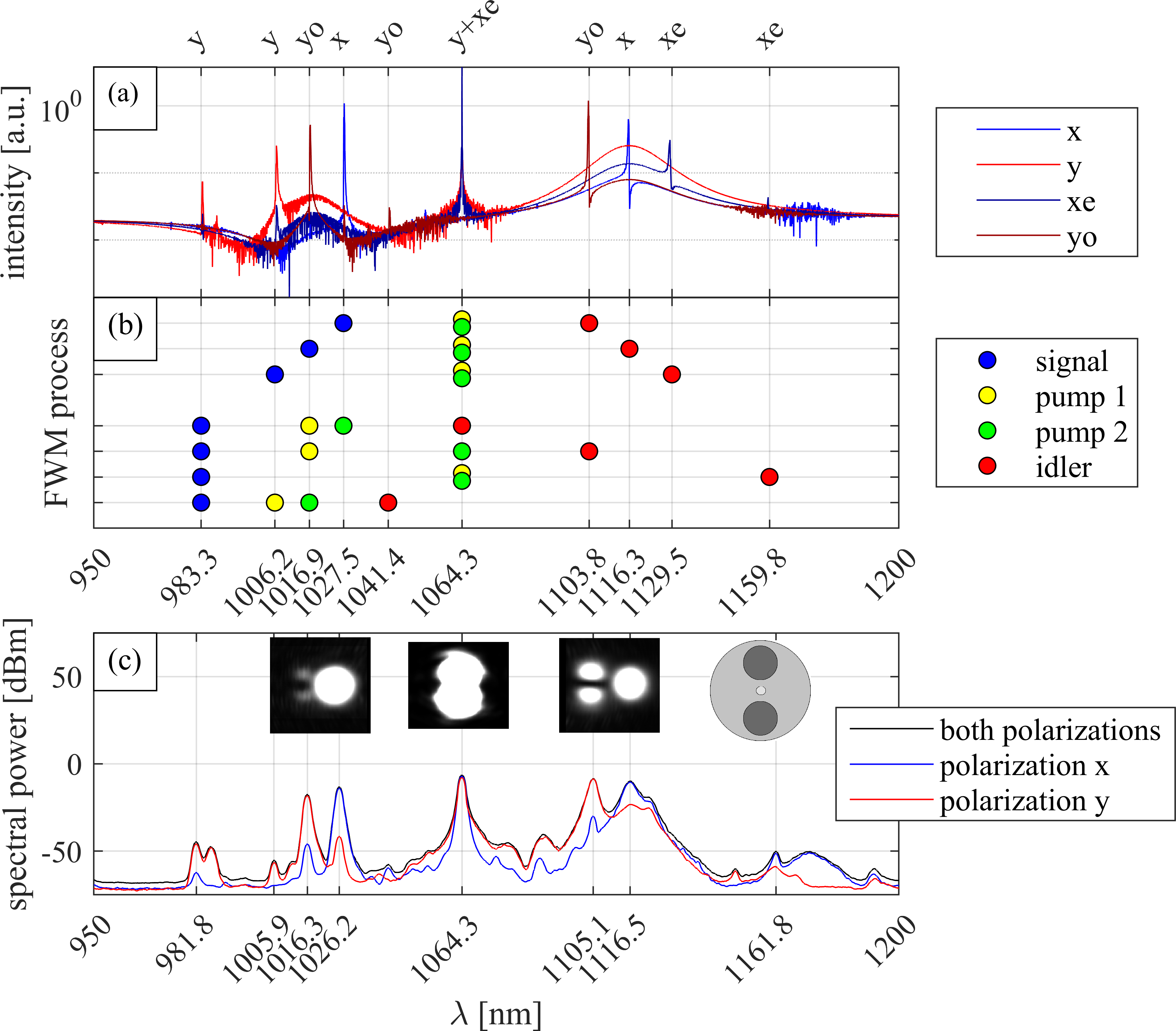}
\caption{(a) Mode resolved spectra obtained in simulations for CW excitation of $y$ and $xe$ modes at \SI{1064.3}{\nano\meter}. (b) Diagram indicating the spectral positions of the bands in distinct spontaneous (top three) and cascaded (bottom four) FWMs. (c)~Polarization-resolved spectra observed experimentally. The insets show images of the pump and generated bands in both polarizations.}
\label{fig:02_IMFWM}
\end{figure}

In the theoretical analysis, FWM conversion was focused upon directly from the pump.
Subsequently, to investigate the cascaded processes, the nonlinear propagation of light was simulated in the considered fiber.
The CNLSE solver was used based on the software implemented by Wright et al.~
\cite{Wright2018}.
Thereafter, the measured properties of the modes, as listed in Table~\ref{tab:fiber properties}, were used to calculate the linear terms in the simulations.
Furthermore, the fiber nonlinearity $n_2 = \SI{2.6e-20}{\meter^2\per\watt}$ and overlap factors, obtained using Eqs.~\ref{eqn:SR}~and~\ref{eqn:SK}, were used to calculate the nonlinear terms.
The simulated spectral positions of the bands generated by the direct FWM were consistent with the theoretical predictions for all excitation cases.
Figure~\ref{fig:02_IMFWM}(a) shows the simulated mode-resolved spectra generated in a \SI{12}{\meter} long piece of the fiber from a CW pump (\SI{1064.3}{\nano\meter}, \SI{1500}{\watt}) coupled equally in the $y$ and $xe$ modes.
Different intermodal-vector FWM processes occur simultaneously for this combination of excited modes.
Primarily, there are: (i) the FWM process that involves only two excited modes, which generates a single signal/idler pair in the same modes; (ii) the two FWM processes that involve four modes, which generate the two signal/idler pairs in the $x$ and $yo$ modes.
Subsequently, there are four cascaded stimulated FWMs. The observed FWM processes are illustrated in Figure~\ref{fig:02_IMFWM}(b).

Finally, experiments were performed to confirm the predictions of the theoretical analysis and numerical simulations.
A \SI{1064.3}{\nano\meter} Nd:YAG laser with a pulse duration of \SI{1}{\nano\second}, repetition rate of \SI{19}{\kilo\hertz} and an average power of \SI{140}{\milli\watt} was employed as the pump. 
The selective excitation of all the considered combinations of modes was experimentally realized using the method described in~\cite{Zolnacz2022}.
Example spectra registered for the excitation of the $y$ and $xe$ modes are shown in Figure~\ref{fig:02_IMFWM}(c), with images of each band captured via a camera placed at the fiber output following a diffraction grating (600 lines/mm).
The measured spectral positions of the signal/idler bands ($\lambda_\mathrm{e}^{(p)}$ and $\lambda_\mathrm{e}^{(n)}$) for all the possible excitations are listed in Table~\ref{tab:FWM spectral}.
However, the position of the single band is missing for the  $(y,xe)\rightarrow(y,xe)$ process because the expected peak is covered by a broad Stokes Raman band.

As shown in Table~\ref{tab:FWM spectral}, the registered spectra generated by spontaneous FWM were consistent with the simulation results for all possible combinations of excited modes.
Furthermore, the measured positions of the bands generated in the cascaded processes were so consistent with the simulations (Fig. 2). 
In addition, the experimental results of this study also prove that in case of excitation by pairs of two spatial modes of orthogonal polarizations, distinct intermodal-vectorial FWM processes can be initiated, resulting in  the generation of multiple signal/idler bands.

In summary, this study presented a variety of FWM processes that are possible between six polarization modes in a birefringent step index fiber.
Subsequently, the role of phase birefringence in the phase-matching condition for intermodal-vectorial FWM processes that occur when two spatial modes of orthogonal polarizations are excited by the pump was highlighted.
The obtained results show that the selective excitation of spatial/polarization modes allows the control of the position of FWM bands in a broad range, which can aid in extending space-divion multiplexing schemes
\cite{Anjum2019}
with polarization degree and in entangled photon pairs sources~\cite{DelaTorre-Robles2021}.

\textbf{Funding}
National Science Centre of Poland (2016/22/A/ST7/00089); National Science Centre of Poland (2018/30/E/ST7/00862).

\bibliographystyle{unsrt}  
\bibliography{sample}

\end{document}